\documentclass[12pt]{iopart}

\usepackage{iopams}  
\usepackage{graphicx}

\begin{document}

\title[Local Path Fitting: A New Approach to Variational Integrators]{Local Path Fitting: A New Approach to Variational Integrators}

\author{D S Vlachos and O T Kosmas}

\address{Department of Computer Science and Technology,
Faculty of Sciences and Technology, University of Peloponnese
GR-22 100 Tripolis, Terma Karaiskaki, GREECE}
\eads{\mailto{dvlachos@uop.gr}, \mailto{odykosm@uop.gr}}
\begin{abstract}
In this work, we present a new approach to the construction of variational integrators. In the general case, the estimation of the action integral in a time interval $[q_k,q_{k+1}]$ is used to construct a symplectic map $(q_k,q_{k+1})\rightarrow (q_{k+1},q_{k+2})$. The basic idea here, is that only the partial derivatives of the estimation of the action integral of the Lagrangian are needed in the general theory. The analytic calculation of these derivatives, give raise to a new integral which depends not on the Lagrangian but on the Euler--Lagrange vector, which in the continuous and exact case vanishes. Since this new integral can only be computed through a numerical method based on some internal grid points, we can locally fit the exact curve by demanding the Euler--Lagrange vector to vanish at these grid points. Thus the integral vanishes, and the process dramatically simplifies the calculation of high order approximations. The new technique is tested for high order solutions in the two-body problem with high eccentricity (up to 0.99) and in the outer solar system.
\end{abstract}

\maketitle

\section{Introduction}
\label{section_intro}
It is well known that the dynamics of seemingly unrelated conservative systems in mechanics, physics, biology, and chemistry fit the Hamiltonian formalism (\cite{rabee_PhD_Diss}). Included among these are particle, rigid body, ideal fluid, solid, and plasma dynamics. An important property of the Hamiltonian flow or solution to a Hamiltonian system is that it preserves the Hamiltonian and the symplectic form (see, for example, \cite{marsden_STAM_17_99}). A key consequence of symplecticity is that the Hamiltonian flow is phase-space volume preserving (Liouville¢s theorem). Since analytic expressions for the Hamiltonian flow are rarely available, approximations based on discretization of time are used. A numerical integration method which approximates a Hamiltonian flow is called symplectic if it discretely preserves a symplectic 2-form to within numerical round off (\cite{vogalaere_TR_DM_UND, ruth_IEEETNS_30_2669_83, feng_JCM_4_279_86}) and standard otherwise. By ignoring the Hamiltonian structure, a standard method often introduces spurious dynamics. This effect is excellent illustrated in \cite{chossat_SIADS_4_1140_05} where a spherical pendulum with $D_4$ symmetric perturbation is integrated using implicit Euler and a symplectic method (variational Euler). The computation of a Poincar\'{e} section shows that the standard method, despite the order of magnitude in integration time-step size, artificially corrupts phase space structures by exhibiting a systematic drift in the invariant tori, whereas variational Euler preserves them. Moreover, in systems that are non-integrable, symplectic integrators often perform much better even compared with projection methods, where the calculated values are projected in the manifold defined by the constant symplectic structure. Finally, as it shown in \cite{dullweber_JCP_107_5840_97}, in many body problems, the symplectic integrators perform increasingly better than standard methods as the number of bodies increases.

Symplectic integrators can be derived by a variety of ways including Hamilton-Jacobi theory, symplectic splitting, and variational integration techniques. Early investigators, guided by Hamilton-Jacobi theory, constructed symplectic integrators from generating functions which approximately solve the Hamilton--Jacobi equation (\cite{vogalaere_TR_DM_UND, ruth_IEEETNS_30_2669_83, feng_JCM_4_279_86}). The symplectic splitting technique is based on the property that symplectic integrators form a group, and thus, the composition of symplectic-preserving maps is also symplectic. The idea is to split the Hamiltonian into terms whose flow can be explicitly solved and then compose these individual flows in such a fashion that the composite flow is consistent and convergent with
the Hamiltonian flow being simulated as it is explained in detail in \cite{leimkuhler_CMACM_14_04}. On the other hand, variational integration techniques determine integrators from a discrete Lagrangian and associated discrete variational principle. The discrete Lagrangian can be designed to inherit the symmetry associated with the action of a Lie group, and hence by a discrete Noether¢s theorem, these methods can also preserve momentum invariants (for a discussion of the above statements, see \cite{rabee_PhD_Diss}).  

Variational integration theory derives integrators for mechanical systems from discrete variational principles (\cite{veselov_FAA_22_83_88, mackay_BOOK_SADHS_CP_92, marsden_AN_10_357_01, wendlandt_PD_106_223_97, bridges_JPAMG_39_5287_06}). Variational principles have been succesfully applied to partial differential equations and to stochastic systems as well {see, for example, \cite{munoz_JPAMG_39_L93_06}). In the general theory, discrete analogs of the Lagrangian, Noether¢s theorem, the Euler--Lagrange equations, and the Legendre transform can be easily obtained \cite{jalnapurkar_JPAMG_39_5521_06, lall_JPAMG_39_5509_06, xie_JPAMG_41_085208_08}. Moreover, variational integrators can readily incorporate holonomic constraints (via Lagrange multipliers) and non-conservative effects (via their virtual work) (\cite{marsden_AN_10_357_01, wendlandt_PD_106_223_97}). The algorithms derived from this discrete principle have been successfully tested in infinite and finite-dimensional conservative, dissipative, smooth and non-smooth mechanical systems (see \cite{amore_JPAMG_40_13047_07, yabu_JPAMG_41_275212_08, lew_IFEM_98_04} and references therein). Recently, variational principles have been applied to particle mesh methods \cite{cotter_JPAMG_41_344003_08} and to fractional stochastic optimal control \cite{atanackovic_JPAMG_41_085208_08}. In the general approach, as it will be presented in section \ref{section_DVM}, the variational principle to apply is the estimation of the action integral in a time interval $[t_0,t_1]$ as a smooth function of the edges of the interval $q_0,q_1$. Since any sufficiently smooth and non-degenerate function $S(q_0,q_1)$ generates via
\begin{eqnarray}
 p_0=-\frac{\partial S(q_0,q_1)}{\partial q_0} &,& p_1= \frac{\partial S(q_0,q_1)}{\partial q_1} \nonumber
\end{eqnarray}
a symplectic map $(q_0,p_0)\rightarrow (q_1,p_1)$ (\cite{hairer_Book_GNISPAODE_S_02}), the estimated action integral can be used to develop a discrete analog to the Euler--Lagrange equations.

The accuracy is not the terminus of the application of variational integrators, but rather their ability to discretely preserve essential structure of the continuous system and in computing statistical properties of larger groups of orbits, such as in computing Poincar\'{e} sections or the temperature of a system (see, for example, \cite{lew_IFEM_98_04, lew_IJNME_60_153_04}). On the other hand, high accuracy can be obtained using special designed methods as it is explained in \cite{kosmas_0904_0112v1, kosmas_0903_3370v1}. These methods, although they produce high order estimations of the positions and the momenta, computationally become very heavy as the order increases. This is due to the large number of parameters that have to be calculated in each step as a solution to a non-linear system, depending on the nature of the Lagrangian. The total set of equations, consists of partial derivatives of the action integral and a set of variational equations that determine the internal points, necessary for high order methods.

In this work, we present a new approach to the construction of variational integrators. The basic idea, is that only the partial derivatives of the estimation of the action integral of the Lagrangian are needed in the general theory. The analytic calculation of these derivatives, give raise to a new integral which depends not on the Lagrangian but on the Euler--Lagrange vector, which in the continuous and exact case vanishes. Since this new integral can only be computed through a numerical method based on some internal grid points, we can locally fit the exact curve by demanding the Euler--Lagrange vector to vanish at these grid points. Thus the integral vanishes, and the process dramatically simplifies the calculation of high order approximations.

\section{Discrete Variational Mechanics}
\label{section_DVM}
The well known least action principle of the continuous Lagrange - Hamilton
Dynamics can be used as a guiding principle to derive discrete integrators.
Following the steps of the derivation of Euler-Lagrange equations in the
continuous time Lagrangian dynamics, one can derive the discrete time
Euler-Lagrange equations. For this purpose, one considers positions $q_{0}$ and
$q_{1}$ and a time step $h\in{R}$, in order to replace the parameters of
position $q$ and velocity $\dot{q}$ in the continuous time Lagrangian $L (q,
\dot{q}, t)$. Then, by considering the variable $h$ as a very small (positive)
number, the positions $q_{0}$ and $q_{1}$ could be thought of as being two
points on a curve (trajectory of the mechanical system) at time $h$ apart. Under
these assumptions, the following approximations hold: $$q_{0}\approx q(0)\, ,
\qquad \qquad q_{1}\approx q(h) \, ,$$ and a function $L_{d}(q_{0},q_{1},h)$
could be defined known as a discrete Lagrangian function.

Many authors assume such functions to approximate the action integral along the
curve segment between $q_{0}$ and $q_{1}$, i.e.
\begin{equation}
 L_{d}(q_{0},q_{1},h)=\int_{0}^{h}L(q,\dot{q},t) dt
\end{equation}
Furthermore, one may consider the very simple approximation for this integral
given on the basis of the rectangle rule described in
\cite{marsden_AN_10_357_01}. According to this rule, the integral
$\int_{0}^{T}{Ldt}$ could be approximated by the product of the time-interval
${h}$ times the value of the integrand $L$ obtained with the velocity $\dot{q}$
replaced by the approximation $(q_{1}-q_{0})/h$:  The next step is to consider a
discrete curve defined by the set of points $\{q_{k}\} _{k=0}^{N}$, and
calculate the discrete action along this sequence by summing the discrete
Lagrangian of the form $L_{d}(q_{k},q_{k+1},h)$ defined for each adjacent pair
of points $(q_{k}$, $q_{k+1})$. 

Following the case of the continuous dynamics, we compute variations of this
action sum with the boundary points $q_{0}$ and $q_{N}$ held fixed. Briefly,
discretization of the action functional leads to the concept of an action sum 
\begin{equation}
S_{d}(\gamma_{d})=\sum_{k=1}^{n-1}L_{d}(q_{k-1},q_{k}),
\qquad \gamma_{d}=(q_{0},...,q_{n-1})\in Q^{n}
\end{equation}
where $L_{d} : Q \times Q \rightarrow R$ is an approximation of L called the
discrete Lagrangian. Hence, in the discrete setting the correspondence to the
velocity phase space $TQ$ is $Q \times Q$. An intuitive motivation for this is
that two points close to each other correspond approximately to the same
information as one point and a velocity vector. The discrete Hamilton's
principle states that if $\gamma_{d}$ is a motion of the discrete mechanical
system then it extremizes the action sum, i. e., $\delta S_{d}=0$. By
differentiation and rearranging of the terms and having in mind that both $q_0$
and $q_N$ are fixed, the discrete Euler-Lagrange (DEL) equation is obtained:
\begin{equation}
\label{equ_DEL}
D_{2}L_{d}(q_{k-1},q_{k},h)+D_{1}L_{d}(q_{k},q_{k+1},h)=0
\end{equation}
where the notation $D_{i}L_{d}$ indicates the slot derivative with respect to
the argument of $L_{d}$.

We can define now the map $\Phi:Q\times Q\rightarrow Q\times Q$, where $Q$ is
the space of generalized positions $q$, by which
\begin{equation}
 D_1L_d\circ \Phi+D_2L_d=0
\end{equation}
which means that $\Phi (q_{k-1},q_k)=(q_k,q_{k+1})$. Then, if for each $q\in Q$,
the map $D_1L_d(q,q):T_qQ\rightarrow T^*_qQ$ is invertible, then $D_1L_d:
Q\times Q\rightarrow T^*Q$ is locally invertible and so the discrete flow
defined by the map $\Phi$ is well defined for small enough time steps (see
\cite{kane_JMP_40_7_3353_99} for details). Moreover, if we define the fiber
derivative
\begin{equation}
 FL_d:Q\times Q\rightarrow T^*Q
\end{equation}
and the two-form $\omega$ on $Q\times Q$ by pulling back the canonical two-form
$\Omega_{CAN}=dq^i\wedge dp_i$ from $T^*Q$ to $Q\times Q$:
\begin{equation}
 \omega=FL^*_d(\Omega_{CAN})
\end{equation}
The coordinate expression for $\omega$ is
\begin{equation}
 \omega=\frac{\partial ^2L_d}{\partial q^i_k \partial
q^j_{k+1}}(q_k,q_{k+1})dq^i_k\wedge d^j_{k+1}
\end{equation}
and can be easily proved that the map $\Phi$ preserves the symplectic form
$\omega$ (two different proofs are presented in \cite{marsden_CMP_199_351_98}
and \cite{wendlandt_PD_106_223_97}). Finally, assuming that the discrete
Lagrangian is invariant under the action of a Lie group $G$ on $Q$ and $\xi \in
g$, the Lie algebra of $G$, by analogy with the continuous case, we can define
the discrete momentum map $J_d:Q\times Q\rightarrow g^*$ by
\begin{equation}
 \left<J_d(q_k,q_{k+1}),\xi\right>:=\left<D_aL_d(q_k,q_{k+1},\xi_Q(q_k)\right>
\end{equation}
It can be proved that the map $\Phi$ preserves the momentum map $J_d$
\cite{wendlandt_PD_106_223_97}.

In a position-momentum form the discrete Euler-Lagrange equations
(\ref{equ_DEL}) can be defined by the equations below 
\begin{eqnarray}
\label{equ_DHP}
\nonumber p_{k}&=&-D_{1}L_{d}(q_{k},q_{k+1},h) \\
p_{k+1}&=&D_{2}L_{d}(q_{k},q_{k+1},h)
\end{eqnarray}

\section{Local Path Fitting}
The system in eq.\ref{equ_DHP} can be considered now as a numerical one-step method $(q_k,p_k)\rightarrow (q_{k+1},p_{k+1})$. The level of the accuracy in estimation of the integral
\begin{equation}
 L_d(q_k,q_{k+1},h)\sim \int_{t_k}^{t_{k+1}}L(q,\dot{q},t)dt
\label{equ_Ld}
\end{equation}
fully characterizes the accuracy of the method. But as we can see from eq.\ref{equ_DHP}, only the derivatives of $L_d$ are needed. Thus, consider a parameter $\lambda$. Then
\begin{equation}
 \frac{\partial L_d(q_k,q_{k+1},h)}{\partial \lambda}=L(q(t_{k+1}),\dot{q}(t_{k+1}),h)\frac{\partial h}{\partial \lambda}+\int_{t_k}^{t_{k+1}}\frac{\partial L(q,\dot{q},t)}{\partial \lambda}dt
\end{equation}
But,
\begin{equation}
 \frac{\partial L(q,\dot{q},t)}{\partial \lambda}=\frac{\partial L(q,\dot{q},t)}{\partial q}\frac{\partial q}{\partial \lambda}+\frac{\partial L(q,\dot{q},t)}{\partial \dot{q}}\frac{\partial \dot{q}}{\partial \lambda}
\end{equation}
or, changing the order of derivation in the second term of the right hand part,
\begin{equation}
 \frac{\partial L(q,\dot{q},t)}{\partial \lambda}=\frac{\partial L(q,\dot{q},t)}{\partial q}\frac{\partial q}{\partial \lambda}+\frac{\partial L(q,\dot{q},t)}{\partial \dot{q}}\frac{d\frac{\partial q}{\partial \lambda}}{dt}
\end{equation}
Integrating by parts now, we get
\begin{eqnarray}
 \frac{\partial L(q,\dot{q},t)}{\partial \lambda}=L(q(t_{k+1}),\dot{q}(t_{k+1}),h)\frac{\partial h}{\partial \lambda}+ \nonumber \\
\int_{t_k}^{t_{k+1}}\left(\frac{\partial L(q,\dot{q},t)}{\partial q}-\frac{d}{dt}\left(\frac{\partial L(q,\dot{q},t)}{\partial \dot{q}}\right)\right)\frac{\partial q}{\partial \lambda}dt+\frac{\partial L(q,\dot{q},t)}{\partial \dot{q}}\frac{\partial q}{\partial \lambda}\Big{|}_{t_k}^{t_{k+1}}
\end{eqnarray}
Now, instead of estimating the integral in eq.\ref{equ_Ld}, we only have to estimate
\begin{equation}
 I_0=\int_{t_k}^{t_{k+1}}\left(\frac{\partial L(q,\dot{q},t)}{\partial q}-\frac{d}{dt}\left(\frac{\partial L(q,\dot{q},t)}{\partial \dot{q}}\right)\right)\frac{\partial q}{\partial \lambda}dt
\end{equation}
We can use any quadrature rule that is based on a set of $S+1$ grid points at times $\{t_k+c^jh,j=0,1,\dots ,S\}$. On the other hand, if we demand that the Euler-Lagrange equation
\begin{equation}
 \frac{\partial L(q,\dot{q},t)}{\partial q}-\frac{d}{dt}\left(\frac{\partial L(q,\dot{q},t)}{\partial \dot{q}}\right)=0
\label{equ_EL}
\end{equation}
holds at these grid points, then $I_0=0$ and
\begin{equation}
 \frac{\partial L(q,\dot{q},t)}{\partial \lambda}=L(q(t_{k+1}),\dot{q}(t_{k+1}),h)\frac{\partial h}{\partial \lambda}+\frac{\partial L(q,\dot{q},t)}{\partial \dot{q}}\frac{\partial q}{\partial \lambda}\Big{|}_{t_k}^{t_{k+1}}
\end{equation}
The set of equations eq.\ref{equ_DHP} are now given:
\begin{eqnarray}
 p_k=-\frac{\partial L(q,\dot{q},t)}{\partial \dot{q}}\frac{\partial q}{\partial q_k}\Big{|}_{t_k}^{t_{k+1}} \nonumber \\
 p_{k+1}=\frac{\partial L(q,\dot{q},t)}{\partial \dot{q}}\frac{\partial q}{\partial q_{k+1}}\Big{|}_{t_k}^{t_{k+1}}
\label{equ_varDHP}
\end{eqnarray}
The above set of equations is consistent with the variational principles. Consider the $S+1$ grid points at times $t^j=t_k+c^jh$, with $c^0=0$ and $c^S=1$ the edge points $q_k,q_{k+1}$. For the internal points we have
\begin{equation}
 \frac{\partial L(q,\dot{q},t)}{\partial \dot{q}}\frac{\partial q}{\partial q^j}\Big{|}_{t_k}^{t_{k+1}}=0\;,j=1,2,\ldots ,S-1
\end{equation}
where $q^j$ are the internal points, because the curve is fixed at its endpoints and thus 
\begin{equation}
 \frac{\partial Ld}{\partial q^j}=0
\end{equation}

Let us now work an exact example. Consider the Lagrangian of the harmonic oscillator with unity frequency
\begin{equation}
 L=\frac{1}{2}\dot{q}^2-\frac{1}{2}q^2
\end{equation}
and let $q(t)$ in the interval $[t_k,t_{k+1}]$ is given
\begin{equation}
 q(t)=q_k\left(1-\frac{t}{h}\right)+q_{k+1}\frac{t}{h}+x\frac{t}{h}\left(1-\frac{t}{h}\right)+y\left(\frac{t}{h}\right)^2\left(1-\frac{t}{h}\right)
\end{equation}
where $h=t_{k+1}-t_k$ and $x,y$ are free parameters. We demand now that the Euler--Lagrange equation holds at points $t=t_k$ and $t=t_{k+1}$. Any quadrature now for the calculation of the action integral based only on edge points will give the set of equations eq.\ref{equ_varDHP}. The parameters $x,y$ are easily calculated
\begin{eqnarray}
 x=\frac{h^2}{3}\left(q_k+\frac{q_{k+1}}{2}\right) \nonumber \\
y=\frac{h^2}{6}\left(q_{k+1}-q_k\right) \nonumber
\end{eqnarray}
Then, the method eq.\ref{equ_varDHP} gives
\begin{eqnarray}
 q_{k+1}=\frac{6h\dot{q}_k+q_k\left(6-2h^2\right)}{6+h^2} \nonumber \\
\dot{q}_{k+1}=\frac{q_{k+1}-q_k}{h}-\frac{h}{3}\left(q_k+\frac{q_{k+1}}{2}\right) \nonumber
\end{eqnarray}
in fig. \ref{fig_res1}, the calculated positions are plotted for the first $10$ periods using a step size $h=0.01$.

\begin{figure}
\center
\includegraphics[scale=0.3]{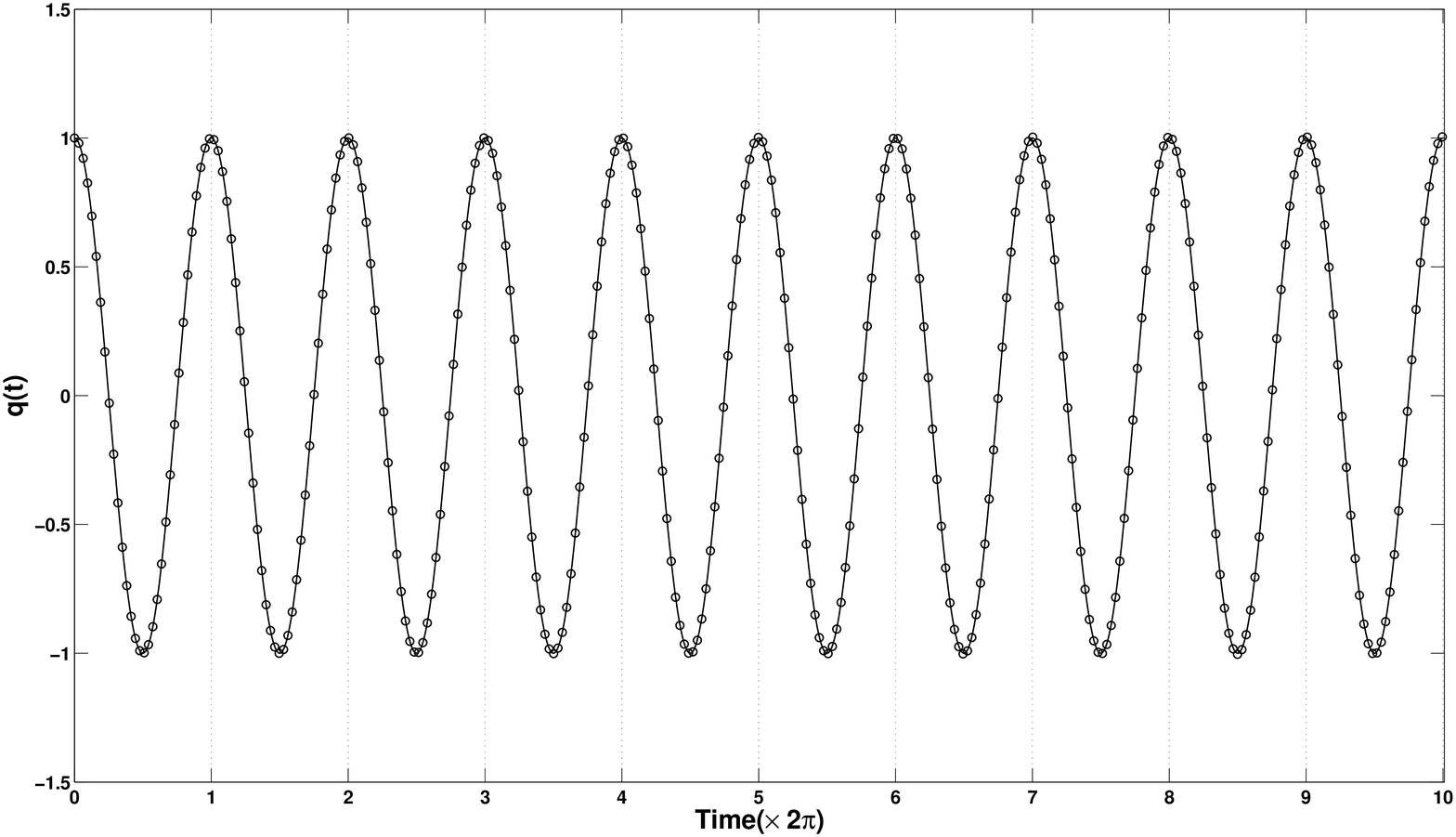}
\caption{The exact orbit (solid line) for the harmonic oscillator with unity frequency and the calculated points ($\circ$) for the first $10$ periods using an orbit satisfying the Euler--Lagrange equation at the edge points.}
\label{fig_res1}
\end{figure}

\section{Local Path fitting using Bernstein basis polynomials}
The $n + 1$ Bernstein basis polynomials of degree $n$ are defined as
\begin{equation}
\label{equ_def_bernstein}
b_{j,n}(x)=\left(
\begin{array}{l}n\\j\end{array}\right)x^j(1-x)^{n-j}\;,\;j=0,1,\ldots ,n
\end{equation}
where the binomial coefficient
\begin{equation}
\left( \begin{array}{l}n\\j\end{array}\right)=\frac{n!}{j!(n-j)!} \nonumber
\end{equation}
The Bernstein basis polynomials of degree $n$ form a basis for the vector space
$\Pi _n$ of polynomials of degree $n$. Some properties of the Bernstein basis polynomials are summarized in the following:
\begin{itemize}
 \item $b_{j,n}(x)=0$ if $j<0$ or $j>n$.
\item $b_{j,n}(0)=\delta _{j,0}$ and $b_{j,n}(1)=\delta _{j,n}$, where $\delta$ is the Kronecker function.
\item $b_{j,n}(1-x)=b_{n-j,n}(x)$
\item $b'_{j,n}(x)=n\left(b_{j-1,n-1}(x)-b_{j,n-1}(x)\right)$
\item The Bernstein basis polynomials of degree $n$ form a partition of unity, i.e. $\sum_{j=0}^nb_{j,n}(x)=1$
\item A Bernstein polynomial can always be written as a linear combination of polynomials of higher degree, i.e. 

$b_{j,n-1}(x)=\frac{n-j}{n}b_{j,n}(x)+\frac{j+1}{n}b_{j+1,n}(x)$.
\end{itemize}

The most important property of Bernstein basis is their capability to approximate continuous functions. Let $f(x)$ be a continuous function on the interval $[0,1]$. If we consider now the Bernstein polynomial
\begin{equation}
 B_n(f)(x)=\sum_{j=0}^n f\left( \frac{j}{n}\right)b_{j,n}(x)
\end{equation}
then
\begin{equation}
\lim_{n\rightarrow \infty} B_n(f)(x)=f(x)
\end{equation}
uniformly on the interval $[0,1]$. To prove this, suppose $K$ is a random variable distributed as the number of successes in $n$ independent Bernoulli trials with probability $x$ of success on each trial; in other words, $K$ has a binomial distribution with parameters $n$ and $x$. Then we have the expected value $E(K/n) = x$. Applying the weak law of large numbers of probability theory
\begin{equation}
 \lim_{n\to\infty}P\left(\left|\frac{K}{n}-x\right|>\delta\right)=0
\end{equation}
for every $\delta > 0$. Because $f$, being continuous on a closed bounded interval, must be uniformly continuous on that interval, we can infer a statement of the form
\begin{equation}
 \lim_{n\rightarrow\infty} P\left(\left|f\left( \frac{K}{n}\right) - f(x) \right| > \varepsilon\right) = 0.
\end{equation}
Consequently
\begin{eqnarray}
    \lim_{n\rightarrow\infty} P\left( \left| f \left( \frac{K}{n} \right) - E\left( f \left( \frac{K}{n} \right) \right)\right| + \left| E\left(f\left(\frac{K}{n}\right)\right)-f(x) \right| > \varepsilon\right)=0 \nonumber \\
    \lim_{n\rightarrow\infty} P\left(\left|f\left( \frac{K}{n}\right) - E\left( f\left( \frac{K}{n}\right) \right) \right| > \varepsilon/2\right) + \nonumber \\
    P\left( \left| E\left( f\left( \frac{K}{n} \right) \right) - f(x)\right| > \varepsilon/2\right) = 0 \nonumber
\end{eqnarray}
And so the second probability above approaches $0$ as $n$ grows. But the second probability is either $0$ or $1$, since the only thing that is random is $K$, and that appears within the scope of the expectation operator $E$. Finally, observe that $E(f(K/n))$ is just the Bernstein polynomial $B_n(f)(x)$.

A more general statement for a function with continuous $k$-th derivative is
\begin{equation}
\| B_n(f)^{(k)} \|_\infty \le \frac{(n)_k}{n^k} \| f^{(k)} \|_\infty \; and \; \|f^{(k)}- B_n(f)^{(k)} \|_\infty \rightarrow 0
\end{equation}
where additionally $\frac{(n)_k}{n^k}= \left(1-\frac{0}{n}\right)\left(1-\frac{1}{n}\right) \cdots \left(1-\frac{k-1}{n}\right)$ is an eigenvalue of $B_n$ and the corresponding eigenfunction is a polynomial of degree $k$.

Consider now the Lagrangian $L(q,\dot{q},t)$ of a given system and the state vector $(q_k,p_k)$ at a given time $t_k$. Let $x^j,j=0,1,\ldots ,S$ a set of $S+1$ parameters and the polynomial
\begin{equation}
 q(t)=\sum_{j=0}^Sx^j b_{j,S}\left(\frac{t-t_k}{h}\right)
\end{equation}
which estimates the position at the interval $[t_k,t_{k+1}]$ with $h=t_{k+1}-t_k$. Since $q(t_k)=x^0$ and $q(t_{k+1})=x^S$, we have
\begin{equation}
 q(t)=q_kb_{0,S}\left(\frac{t-t_k}{h}\right)+\sum_{j=1}^{S-1}x^j b_{j,S}\left(\frac{t-t_k}{h}\right)+q_{k+1}b_{S,S}\left(\frac{t-t_k}{h}\right)
\end{equation}
and 
\begin{eqnarray}
 \dot{q}(t)=\sum_{j=1}^{S-1}x^j\frac{S}{h}\left( b_{j-1,S-1}\left(\frac{t-t_k}{h}\right)-b_{j,S-1}\left(\frac{t-t_k}{h}\right)\right)+\nonumber \\
q_{k+1}\frac{S}{h}\left(b_{S-1,S-1}\left(\frac{t-t_k}{h}\right)-b_{S-1,S}\left(\frac{t-t_k}{h}\right)\right)
\end{eqnarray}
Replacing now the position and velocities in the Lagrangian and demanding the Euler--Lagrange equation to hold in grid points, we get

\begin{eqnarray}
 p_k=-\frac{\partial L}{\partial \dot{q}}\Big{|}_{t=t_k} \nonumber \\
\frac{\partial L}{\partial q}\Big{|}_{t=t_k+c^jh}-\frac{d}{dt}\left(\frac{\partial L}{\partial \dot{q}}\right)\Big{|}_{t=t_k+c^jh}=0\;,\;j=1,2,\ldots ,S-1 \nonumber \\
p_{k+1}=\frac{\partial L}{\partial \dot{q}}\Big{|}_{t=t_{k+1}}
\end{eqnarray}
The above system is solved for $(x^j,p_{k+1}),,j=1,2,\ldots ,S$, and gives the next point $q_{k+1}=x^S$.

\section{Numerical Tests}
\subsection{The 2-body problem}
We now turn to the study of two objects interacting through a central force. The
most famous example of this type, is the Kepler
problem (also called the two-body problem) that describes the motion of two
bodies which attract each other. In the solar system the gravitational
interaction between two bodies leads to the elliptic orbits of planets and the
hyperbolic orbits of comets.

If we choose one of the bodies as the center of our coordinate system, the
motion will stay in a plane. Denoting the position of
the second body by $\textbf{q}=(q_{1},q_{2})^{T}$, the Lagrangian of the system
takes the form (assuming masses and gravitational constant equal to 1)
\begin{equation}
L(\textbf{q},\dot{\textbf{q}},t)=\frac{1}{2}\dot{\textbf{q}}^T
\dot{\textbf{q}}+\frac{1}{|\textbf{q}|}
\end{equation}
The initial conditions are taken
\begin{equation}
 \textbf{q}=(1-\epsilon,0)^T\;,\;\dot{\textbf{q}}=\left(0,\sqrt{\frac{1+\epsilon
}{1-\epsilon}}\right)^T
\end{equation}
where $\epsilon$ is the eccentricity of the orbit. In the first experiment, we take the eccentricity $\epsilon =0.5$ and step size $h=0.05$ and plot the relative error in the energy during one period for several number of intermediate points $S$. The results are shown in fig. \ref{fig_res2} where it is clear that the order of the approximation is increased with increasing number of intermediate points.
\begin{figure}
\center
\includegraphics[scale=0.3]{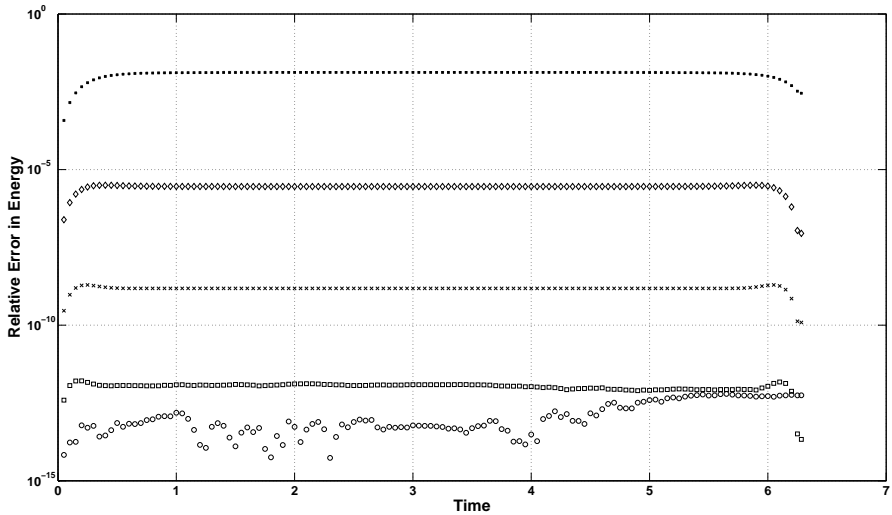}
\caption{The relative error in energy during one period for eccentricity $\epsilon =0.5$, step size $h=0.05$ and for different values of $S$ (number of intermediate points). $(\blacksquare)$ for $S=3$, $(\Diamond )$ for $S=5$, $(\times )$ for $S=7$, $(\Box )$ for $S=9$ and $(\circ )$ $S=11$.}
\label{fig_res2}
\end{figure}

In the second experiment, we take the eccentricity $\epsilon =0.99$ and use an adaptive time step control in order to keep the relative error in energy smaller than $10^{-7}$. Table \ref{table_res1} shows the number of integration steps needed for one period. The order of approximation is again clearly increases with increasing $S$ as the mean step size needed for the same error in energy increases from $1.8\cdot 10^{-3}$ for $S=5$ to $0.1$ for $S=12$.
\begin{table}[ht]
\caption{Number of integration steps}
\centering
\begin{tabular}{cc}
 \hline\hline \textbf{S} &  \textbf{No of Steps} \\
 \hline 
  3 &  $>10^4$ \\
  4 &  $>10^4$  \\
  5 &  3526  \\
  6 &  460   \\
  7 &  421 \\
  8 &  181  \\
  9 &  142  \\
  10 &  112   \\
  11 &  98  \\
  12 &  59   \\
 \hline
\end{tabular}
\label{table_res1}
\end{table}
Finally, in the last experiment we integrate the two-body problem with eccentricity $\epsilon =0.99$ for $10^4$ periods in order to check the long term behavior of the method. The number of intermediate points is $S=12$ and the step size is adaptively controlled in order to keep the relative error in energy less than $10^{-7}$. Fig. \ref{fig_res3} shows the results. In the upper sub-figure, we plot the position during the last period along with the exact solution while in the other, the relative error in angular momentum.

\begin{figure}
\center
\includegraphics[scale=0.3]{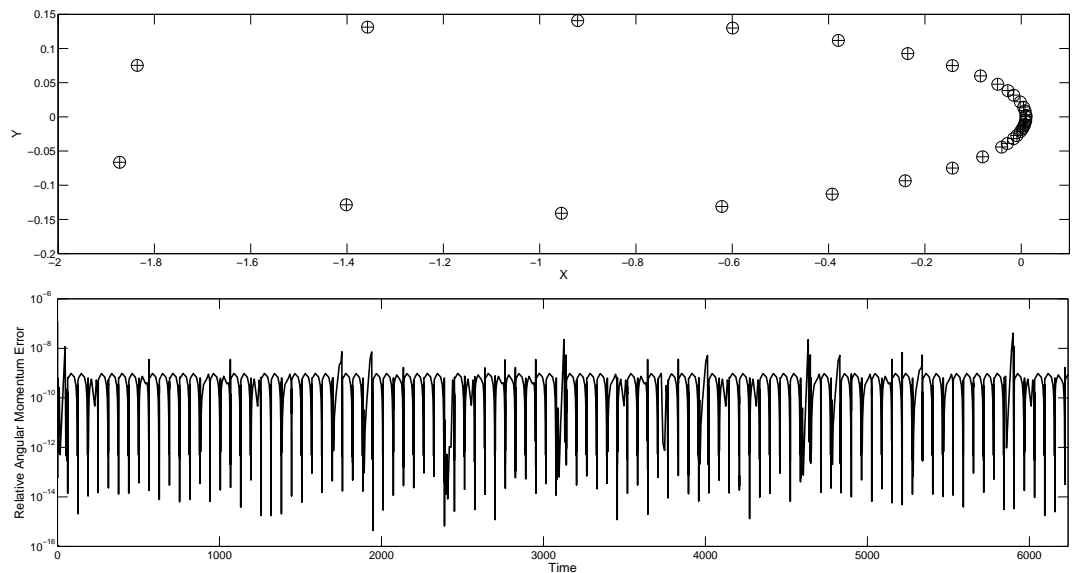}
\caption{Long term integration of the 2-body problem with eccentricity $\epsilon =0.99$ for $10^3$ periods with $S=12$ intermediate points and a step size which is adaptively calculated in order to keep the relative error in energy less than $10^{-7}$. In the first plot, the calculated positions $(+)$ and the exact ones $(\circ)$ are plotted for the last period. In the second plot, the relative error in angular momentum is plotted.}
\label{fig_res3}
\end{figure}

\subsection{The 5-Outer Planet System}
The next problem concerns the motion of the five outer planets relative to the sun. The problem falls in the category of the N-Body problem which is the problem that concerns the movement of N bodies under Newton's law of gravity. The Lagrangian of this system is
\begin{equation}
L(q^i,\dot{q}^i,t)=\frac{1}{2}\sum_{j=0}^{5}m^i\left(\dot{q}^i\right)^T\dot{q}^i+G\sum_{j=1}^{5}\sum_{k=0}^{j-1}\frac{m^j\cdot m^k}{\| q^j-q^k\|}
\label{equ_N_body}\end{equation}
where $G$ is the gravitational constant, $m^j$ is the mass of body $j$ and $\mathbf{q^i},\mathbf{\dot{q}^i}$ are the vectors of the position and velocity of body $i$. In \cite{hairer_Book_GNISPAODE_S_02} the data for the five outer planet problem is given (these data are summarized in table \ref{table_5op}). Masses are relative to the sun, so that the sun has mass 1. In the computations the sun with the four inner planets are considered one body, so the mass is larger than one. Distances are in astronomical units, time is in earth days and the gravitational constant is $G = 2.95912208286 \cdot 10^{-4}$. The Lagrangian (\ref{equ_N_body}) has been integrated for $t \in [0, 10^6 ]$ with step size $h=50$ days and using $S=6$ intermediate points. 
The results are shown in fig. \ref{fig_res4}. The planet orbits are stable, the maximum relative error in energy is $\sim 10^{-7}$, the maximum relative error in momentum is less than $10^{-10}$ and finally the relative error in angular momentum is $\sim 10^{-9}$. Note here, that errors in angular momentum are caused by the round off error produced by the solution of the non-linear system used to calculate the intermediate points. 
\begin{figure}
\center
\includegraphics[scale=0.3]{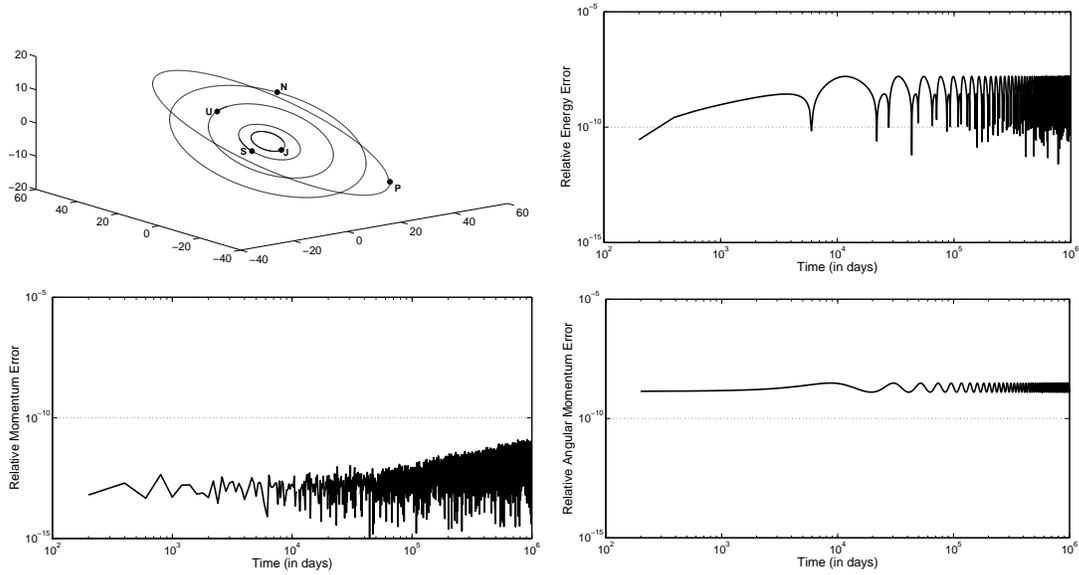}
\caption{Integration of the outer solar system for $10^6$ days with step size $h=50$ days and $S=6$ intermediate points. The planets (J-upiter, S-aturn, U-ranus, N-eptune and P-luto) follow constant orbits, while the relative error in energy is $\sim 10^{-7}$, the maximum relative error in momentum is less than $10^{-10}$ and finally the relative error in angular momentum is $\sim 10^{-9}$.}
\label{fig_res4}
\end{figure}

\begin{table}[ht] 
\caption{Initial Data for the 5-outer planet problem}
\centering      
\begin{tabular}{c | c | c | c}  
\hline
\hline
Planet & Mass & Initial Position & Initial Velocity \\
\hline
Sun & 1.00000597682 & 0 & 0\\
 &  & 0 & 0\\
 &  & 0 & 0\\
\hline
Jupiter & 0.000954786104043 & -3.5023653 & 0.00565429\\
 &  & -3.8169847 & -0.00412490\\
 &  & -1.5507963 & -0.00190589\\
\hline
Saturn & 0.000285583733151 & 9.0755314 & 0.00168318\\
  & & -3.0458353 & 0.00483525\\
 &  & -1.6483708 & 0.00192462\\
\hline
Uranus & 0.0000437273164546 & 8.3101420 & 0.00354178\\
  & & -16.2901086 & 0.00137102\\
  & & -7.2521278 & 0.00055029\\
\hline
Neptune & 0.0000517759138449 & 11.4707666 & 0.00288930\\
  & & -25.7294829 & 0.00114527\\
 &  & -10.8169456 & 0.00039677\\
\hline
Pluto & $1/(1.3 \cdot 10^8 )$ & -15.5387357 & 0.00276725\\
 &  & -25.2225594 & -0.00170702\\
 &  & -3.1902382 & -0.00136504 \\
\hline \hline
\end{tabular} 
\label{table_5op}  
\end{table}

\section{Conclusions}
A new approach to the construction of variational integrators has been developed in this work. The new technique is based on the fact that in order to construct the symplectic map in the variational integrator, we need only the partial derivatives of the estimation of the action integral in a time interval. These derivatives are now functions of the integral of the Euler--Lagrange vector, which in the exact case, vanishes. Thus, taking  an orbit which satisfies the Euler--Lagrange equation in a number of grid points, any quadrature used for the calculation of the action integral vanishes and the process is dramatically simplified. Experimental tests show that these methods are efficiently integrate stiff systems (like the 2-body problem with eccentricity up to $0.99$) conserving all the benefits of the classical variational integrators.

\ack
This paper is part of the 03ED51 research project, implemented within the framework of the "\emph{Reinforcement Programme of Human Research Manpower}" (\textbf{PENED}) and co-financed by National and Community Funds (25\% from the Greek Ministry of Development-General Secretariat of Research and Technology and 75\% from E.U.-European Social Fund).

\section*{References}
\bibliographystyle{unsrt}

\end{document}